\documentclass[11pt,twoside]{article}
\usepackage{asp2014}

\aspSuppressVolSlug
\resetcounters

\begin{document}

\title{J-PLUS Tracking Tool: Scheduler and Tracking software for the Observatorio Astrof{\'i}sico de Javalambre (OAJ)}

\author{Tamara~Civera$^1$}
\affil{$^1$Centro de Estudios de F{\'i}sica del Cosmos de Arag{\'o}n (CEFCA), Teruel, Arag{\'o}n, Spain; \email{tcivera@cefca.es}}

\paperauthor{Tamara~Civera}{tcivera@cefca.es}{0000-0002-0358-8503}{Centro de Estudios de F{\'i}sica del Cosmos de Arag{\'o}n (CEFCA)}{UPAD}{Teruel}{Arag{\'o}n}{44001}{Spain}




\begin{abstract}

The Javalambre Photometric Local Universe Survey (J-PLUS) is an ongoing 12 band photometric optical survey, observing thousands of square degrees of the Northern Hemisphere from the dedicated JAST80 telescope at the Observatorio Astrof{\'i}sico de Javalambre (OAJ). Observational strategy is a critical point in this large survey.

To plan the best observations, it is necessary to select pointings depending on object visibility, the pointing priority and status and location and phase of the Moon. In this context, the J-PLUS Tracking Tool, a web application, has been implemented, which includes tools to plan the best observations, as well as tools to create the command files for the telescope; to track the observations; and to know the status of the survey.

In this environment, robustness is an important point. To obtain it, a feedback software system has been implemented. This software automatically decides and marks which observations are valid or which must be repeated. It bases its decision on the data obtained from the data management pipeline database using a complex system of pointing and filter statuses.

This contribution presents J-PLUS Tracking Tool and all feedback software system.
  
\end{abstract}

\section{Introduction}

The Observatorio Astrof{\'i}sico de Javalambre (OAJ\footnote{\url{https://oaj.cefca.es/}}) is a Spanish astronomical facility particularly conceived for carrying out large sky surveys with two large field of view telescopes: the JST250, a 2.5m telescope of 3deg field of view, and the JAST80, an 80cm telescope of 2deg field of view. The first of these surveys is the Javalambre Photometric Local Universe Survey (J-PLUS \footnote{\url{https://j-plus.es/}}; \citet{2019A&A...622A.176C}), an unprecedented and ongoing photometric sky survey, observing 8500 square degrees of the Northern Hemisphere with JAST80 telescope using a unique set of 12 broad, intermediate and narrow band filters.

In this large survey, observational strategy is a critical point. The characteristics and strategy of the survey are optimized to extract the maximum scientific return in different areas of Astrophysics. In J-PLUS, each pointing is observed in the 12 different filters, and in each filter in 3 consecutive exposures. Moreover, the exposure times are computed automatically in advance according to the phase and distance of the Moon to guarantee homogeneity in the final depths and quality of the photometry.

\section{J-PLUS Tracking Tool}

J-PLUS Tracking Tool is a web portal implemented to deal with the complexity of carrying out this large area survey with thousands of pointings in different filters securing the quality of the data. It includes tools to plan the best observations during the night; to create the command files for the telescope; to track the observations; and to know the status of the survey.

\articlefigure[width=.8\textwidth]{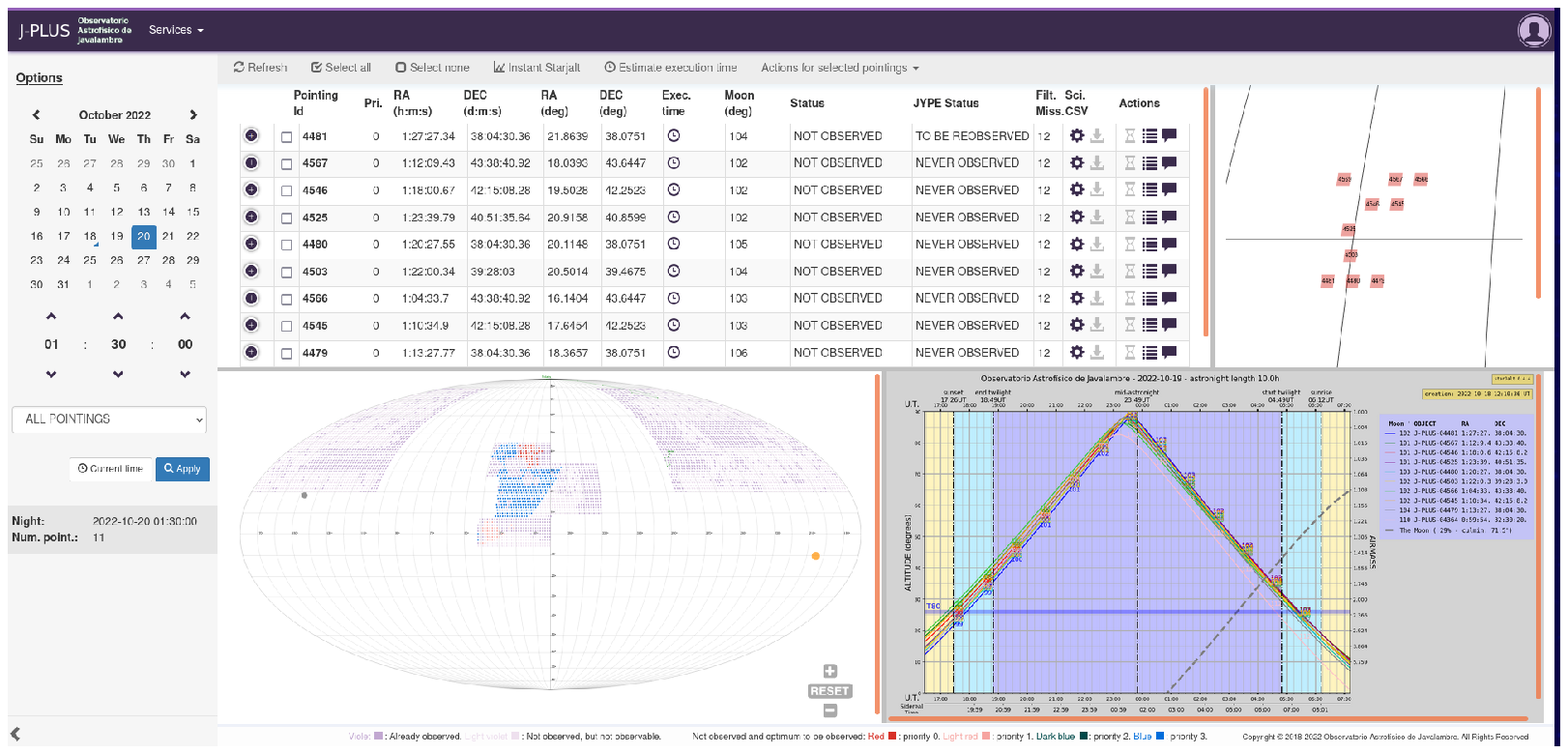}{ex_fig1}{J-PLUS Tracking Tool portal user interface}

\subsection{Night planning tool}

The night planning tool obtains the best observations for the night depending on object visibility, the pointing priority and status and the location and phase of the Moon. More precisely, it calculates them getting all the pointings visible that must be completely or partially (re)observed, ordering them by priority and visibility and discarding those that do not apply with Moon distance survey requirements. Each pointing has an assigned priority from 0 to 3 (0 more priority) depending on the survey scientific objectives.

\subsection{Creating the command files for the telescope}

J-PLUS Tracking Tool also has the functionality to generate a CSV command file understandable by the JAST80 telescope (\citet{10.1117/12.2313208}) to observe the pointing indicated. This command file is generated filling only the instructions to observe the pointing in the filters needed (all or only a set of them), and automatically computing each exposure time according to the filter characteristics and the phase and distance of the Moon at this moment.

\subsection{Survey status viewer}

The survey status viewer allows visualizing all the survey pointings and their status, displaying also statistics of the survey progress by pointings priorities and statuses.

\subsection{Tracking the observations}

The planning of an observing night requires the knowledge of the fields observed, the quality of the data already secured, and the ones still to be observed to optimize scientific returns. So, a good tracking system is necessary.

J-PLUS Tracking Tool gives telescope operators a simple and easy user interface to track the observations where they can manually change the general status of a pointing during the night operation. The different pointings status they can manage are: `not observed' (the pointing must be observed or reobserved), `CSV generated' (the command file for the telescope has been generated and sent to the telescope) and `observed to be validated' (the pointing has been observed, but images must be still checked) (see Figure \ref{ex_fig2}). 

\articlefiguretwo{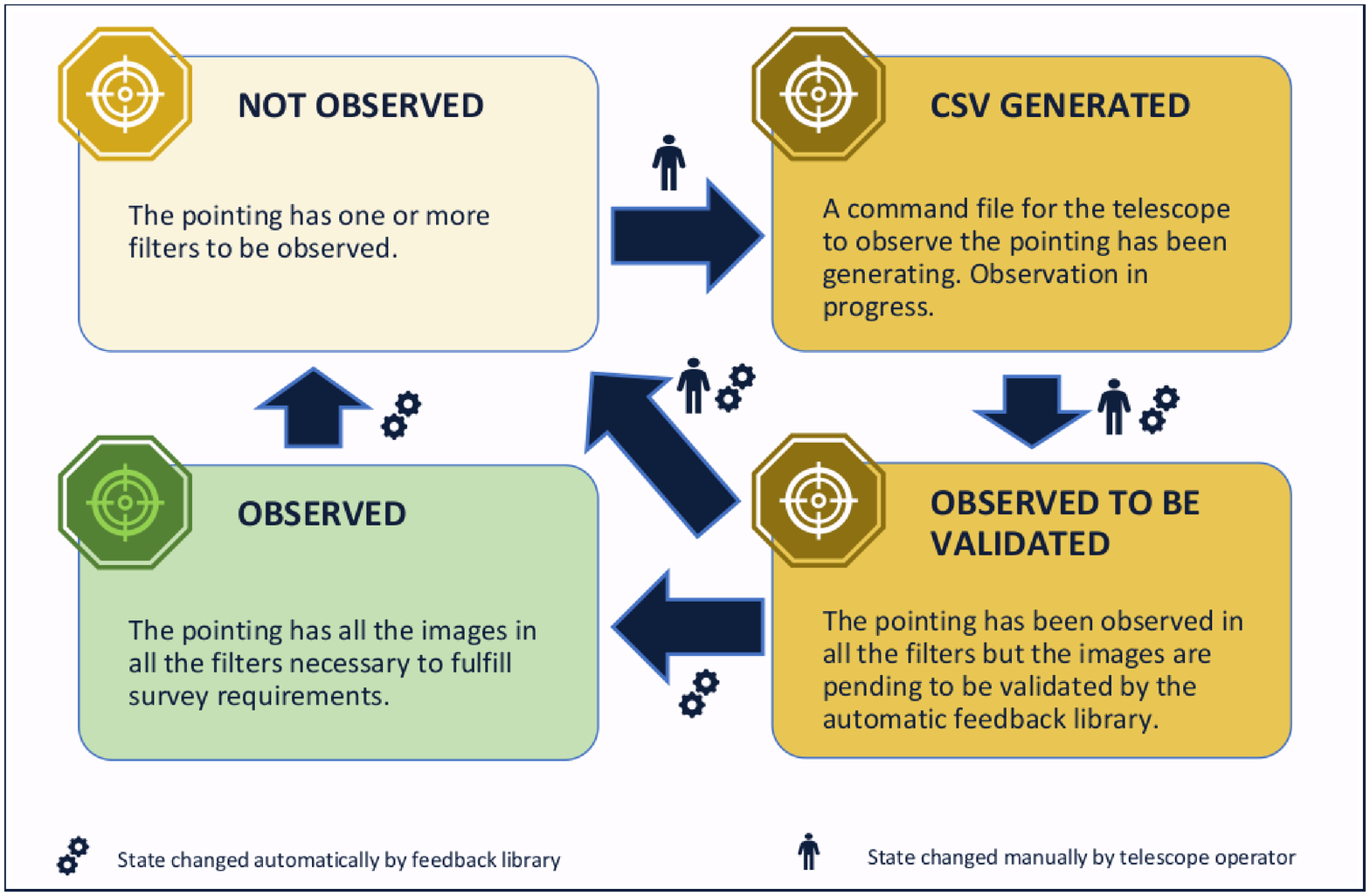}{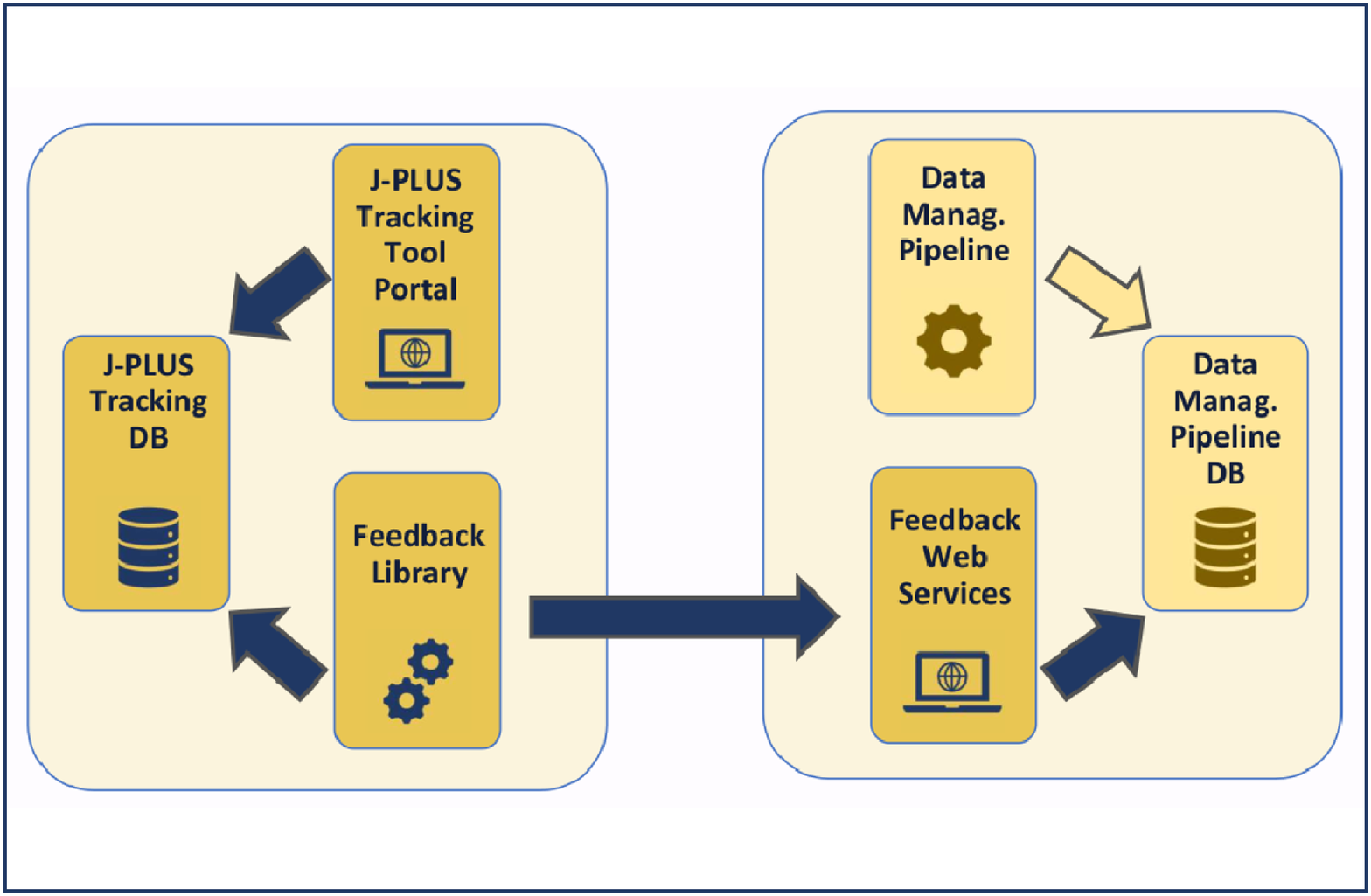}{ex_fig2}{\emph{Left:} Pointings general statuses.  \emph{Right:} J-PLUS Tracking Tool and feedback system architecture.}

To give greater robustness to observations tracking, a complex feedback software system has been designed and implemented. This system obtains everyday different data of the raw and reduced survey images from the data management pipeline database (\citet{2014SPIE.9152E..0OC}) to automatically decide and mark which observations are valid or which must be completely or partially repeated because they have problems or not fulfil J-PLUS survey requirements. This feedback software bases its decisions in a complex pointing and filter statuses system.

\section{The tracking feedback software system}

The feedback software system is composed by the feedback web services and the feedback library (see Figure \ref{ex_fig2}). Feedback web services obtain data of the raw and reduced survey images from the data management pipeline database, translating it in pointing and filter statuses (see Table \ref{table:filterPointingsStatuses}). While the feedback library asks them to automatically take the decision, based on these statuses, if a pointing is valid or it must be partially, or completely reobserved and modifies the general pointing status.

In general lines, if the pointing status given by the system is `Never observed', `To be reobs' or  `Incomplete', the feedback library sets the general status of the pointing as `Observed' so the pointing is set as a candidate to be eligible by the night planning tool. If the pointing status is `Obs pending', the general status is set to `Observed to be validated'. And, finally, if it is `Obs validated', the general status is set to `Observed', which indicates that the pointing has been completely observed. This last general status can only be set by this feedback library.

\begin{table}[!ht]
\caption{Pointings and filter statuses obtained from data management pipeline db}
\smallskip
\begin{center}
{\small
\begin{tabular}{llc}  
\tableline
\noalign{\smallskip}
&\textbf{FILTER STATUSES}\\
\tableline
\noalign{\smallskip}
NEVER OBSERVED & The pointing has been never observed in that filter.\\
\noalign{\smallskip}
\tableline
\noalign{\smallskip}
RAW NO OK & The pointing has been observed in that filter but there are not 3 or\\
& more raw images OK. The filter must be reobserved.\\
\noalign{\smallskip}
\tableline
\noalign{\smallskip}
RAW OK NO PROC & Raw images are OK, but the images have not been already \\
& reduced. It is necessary to wait.\\
\noalign{\smallskip}
\tableline
\noalign{\smallskip}
PROC NO OK & Raw images are OK, but the processed ones are not OK. \\
& The filter must be reobserved.\\
\noalign{\smallskip}
\tableline
\noalign{\smallskip}
PROC OK & Raw and processed images are OK, but survey requirements are \\
& not still been checked. It is necessary to wait.\\
\noalign{\smallskip}
\tableline
\noalign{\smallskip}
PRJ REQ NO OK & Raw and processed images are OK but they do not fulfill \\
&  survey requirements. The filter must be reobserved.\\
\noalign{\smallskip}
\tableline
\noalign{\smallskip}
NOT NUM FILTERS & Raw and processed images are OK, but the pointing has few filters \\
& OK so all filters should be reobserved.\\
\noalign{\smallskip}
\tableline
\noalign{\smallskip}
OBS VALIDATED & Raw images are OK, processed images are OK and they fulfill \\
& survey requirements. The filter images are valid.\\
\noalign{\smallskip}

\tableline
\noalign{\smallskip}
& \textbf{POINTING STATUSES}\\
\noalign{\smallskip}

\tableline
\noalign{\smallskip}
NEVER OBSERVED & The pointing has been never observed.\\
\noalign{\smallskip}
\tableline
\noalign{\smallskip}
TO BE REOBS & There are some images of the pointing already observed, but that \\
&pointing must be completely reobserved.\\
\noalign{\smallskip}
\tableline
\noalign{\smallskip}
INCOMPLETE & It is necessary to only (re)observed some pointing filters.\\
\noalign{\smallskip}
\tableline
\noalign{\smallskip}
OBS PENDING & All filter images seen to be OK, but information still pending to set \\
& it as valid.\\
\noalign{\smallskip}
\tableline
\noalign{\smallskip}
OBS VALIDATED & The pointing has been completely observed and the images fulfill \\
 & survey requirements in all the filters.\\
\noalign{\smallskip}
\label{table:filterPointingsStatuses}
\end{tabular}
}
\end{center}
\end{table}

\section{Conclusions}

J-PLUS Tracking Tool and the different tools offered by it have been presented as well as the feedback software system, a software designed and implemented to improve observations tracking robustness.


\acknowledgements Funding: Fondo de Inversiones de Teruel, 
\\ PID2021-124918NB-C41, PID2021-124918NB-C42

\bibliography{P50}


\end{document}